\documentclass[]{raa}            

\usepackage{graphicx,times}             

\begin{document}

   \title{A Low-mass-ratio and Deep Contact Binary as the Progenitor of the Merger V1309 Sco
}

   \volnopage{Vol.0 (200x) No.0, 000--000}      
   \setcounter{page}{1}          

   \author{Liying Zhu
      \inst{1,2,3}
   \and Ergang Zhao
      \inst{1,2,3}
   \and Xiao Zhou
      \inst{1,2,3}
   }

   \institute{Yunnan Observatories, Chinese
     Academy of Sciences, P.O. Box 110, 650011 Kunming, P. R. China; {\it zhuly@ynao.ac.cn}\\
        \and
        Key laboratory of the structure and evolution of celestial objects, Chinese Academy of Sciences, P.O. Box 110, 650011 Kunming, P. R. China\\
        \and
        University of the Chinese Academy of Sciences, Yuquan Road 19\#, Sijingshang Block, 100049 Beijing, P. R. China\\
   }

   \date{Received~~2009 month day; accepted~~2009~~month day}

\abstract{ Nova Sco 2008 (=V1309 Sco) is one of the V838 Mon type eruptions
rather than a typical classical nova. This enigmatic object was
recently shown to have resulted from the merger of the two stars in
a contact binary. It is the first stellar merger that was undergoing
a common envelope transient. To understand the properties of its
binary progenitor, the pre-outburst light curves were analyzed by
using the W-D (Wilson and Devinney) method. The photometric solution
of the 2002 light curve shows that it is a deep contact binary
($f=89.5(\pm40.5)\,\%$) with a mass ratio of 0.094. The asymmetry of
the light curve is explained by the presence of a dark spot on the
more massive component. The extremely high fill-out factor suggests
that the merging of the contact binary is driven by dynamical mass
loss from the outer Lagrange point. However, the analysis of the
2004 light curve indicates that no solutions were obtained even at
an extremely low mass ratio of q=0.03. This suggests that the common
convective envelope of the binary system disappeared and the
secondary component has spiraled in the envelope of the primary in
2004. Finally, the ejection of the envelope of the primary produced
the outburst.
}

   \authorrunning{L.-Y. Zhu, E.-G. Zhao \& X. Zhou }            
   \titlerunning{The progenitor of the merger V1309 Sco}  

   \maketitle

%
%
\section{Introduction}           

Contact systems are short-period close binary stars where both components are filling their critical roche lobes and sharing a common envelope. They are formed from near-contact binaries via mass transfer and/or angular momentum loss (e.g., Qian 2002a, b; Zhu \& Qian 2006, 2009; Zhu et al. 2009, 2012). This kind binary stars are oscillating around a critical mass ratio (e.g., Qian 2001 a, b; 2003a, b) and will merge into a rapidly rotating single stars (e.g., Qian et al. 2005a; Zhu et al. 2005, 2011). Searching for the mergers of contact binaries is a key question in stellar astrophysics.

V1309 Sco was discovered as Nova Sco 2008 on JD\,2454712 in
September 2008 (Nakano 2008). However, the subsequent evolution
showed that it belongs to a new type of outburst rather than a
typical classical nova (e.g., Mason et al. 2010). Early
spectroscopic data revealed an F-type giant, and then evolved to K-
and M-types (Mason et al. 2010; Rudy et al. 2008a, b). As pointed
out by Tylenda et al. (2011), it shares the principal characteristic
of the V838 Mon type eruptions, i.e evolution to very low effective
temperatures after maximum brightness and during the decline
(Tylenda \& Soker 2006). Some common features of V1309 Sco and the
V838 Mon type outbursts include the outburst amplitude of
7-10\,magnitudes, the eruption time scale of the order of months,
expansion velocities of a few hundreds km/s (instead of a few
thousands as in classical novae), and complete lack of any
high-ionization features.

By using archive photometric data collected in the OGLE project
during about six years before the outburst, Tylenda et al. (2011)
concluded that the progenitor of V1309 Sco was a contact binary with
a period of about 1.4\,days. The binary system quickly evolved
towards its merger and produced the eruption observed in 2008.
Nandez et al. (2014) pointed out that the progenitor consists of a
1.52\,$M_{\odot}$ giant and a 0.16\,$M_{\odot}$ companion with an
orbital of $\sim1.4$\,days and evolves toward the merger primarily
because of Darwin instability. The investigation of Pejcha (2014)
indicated that the period decay timescale $P/\dot{P}$ decreased from
$\sim1000$ to $\sim170$\,years in about six years revealing a
variable rate of mass loss. McCollum et al. (2014) showed the merger
remnant's brightness in optical bandpasses, near-IR bandpasses, and
the Spitzer 3.6\,$\mu$m and 4.5\,$\mu$m channels has varied by
several magnitudes and in complex ways suggesting the occurrence of
a dust formation event. The purpose of the present paper is to
understand the origin of the outburst by analyzing the light curves
of the progenitor of V1309 Sco. Our results indicate that the progenitor of V1309 Sco is a low mass ratio, deep contact binary that is nearly filling the outer critical Roche lobe and undergoing rapidly mass and angular momentum loss.


\section{Analysis of the light curves}

  \begin{figure*}
   \centering
   \includegraphics[width=\textwidth, angle=0]{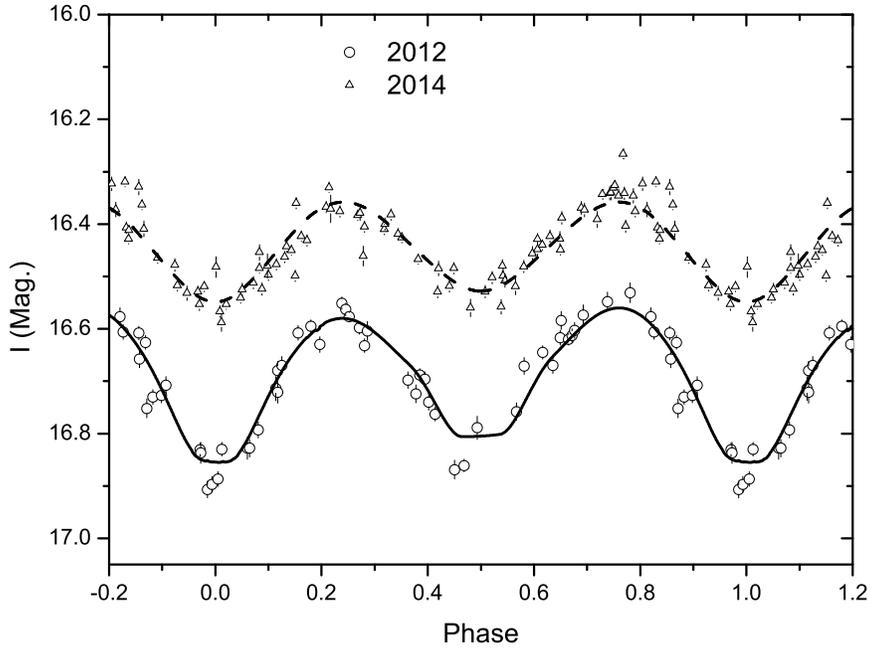}
  \caption{Theoretical
light curves calculated by using the W-D method. Open circles refer
to those I-band data points observed in 2002, while open triangles
to those observations in 2004. The theoretical light curve (the
solid line) obtained by using data observed in 2002 reveals that
V1309 Sco is a low-mass-ratio and deep contact binary, while the one
(the dashed line) for 2004 observations was computed by fixed
q=0.03.}
   \label{fig1}
   \end{figure*}

V1309 Sco was included in one of the stellar fields by the OGLE
team. Therefore, precise photometric data in I-band were obtained
for six years before outburst and were presented by Tylenda et al.
(2011). Those photometric data obtained in 2002 were analyzed by
using the W-D method (Wilson \& Devinney 1971; 1994). The
corresponding light curve is shown in Fig. 1 as open circles. The
temperature for the primary (star 1, the hotter component star
eclipsed at the primary light minimum) was taken as $T_1=4500$\,K
(Tylenda et al., 2011). The bolometric albedo $A_{1}=A_{2}=0.5$ and
the gravity-darkening coefficients $g_{1}=g_{2}=0.32$ were used
because of their convective envelopes. For a detailed treat of limb
darkening, the logarithmic functions for both the bolometric and
bandpass limb-darkening laws were used. The bolometric
limb-darkening coefficients $x_{1bolo}$, $x_{2bolo}$, $y_{1bolo}$,
and $y_{2bolo}$, and the passband-specific limb-darkening
coefficients, $x_{1I}, x_{2I}$, $y_{1I}$, and $y_{2I}$ were chosen
from van Hamme's table (1993) and are listed in Table 1. It is found
that solutions converged at mode 3 and the adjustable parameters
are: the orbital inclination i; the mean temperature of star 2,
$T_{2}$; the monochromatic luminosity of star 1, $L_{1I}$; and the
dimensionless potential ($\Omega_{1}=\Omega_{2}$ for mode 3).

\begin{table}
\caption{Photometric solutions for V1309 Sco.}
\begin{tabular}{lll}
\hline\hline
Parameters & Photometric elements & errors\\
\hline
$g_{1}=g_{2}$&0.32 & assumed \\
$A_{1}=A_{2}$&0.5  & assumed \\
$x_{1bolo}=x_{2bolo}$ & 0.313  & assumed\\
$y_{1bolo}=y_{2bolo}$& 0.660  & assumed\\
$x_{1I}=x_{2I}$ &+0.356  & assumed\\
$y_{1I}=y_{2I}$ &+0.360  & assumed\\
$T_{1}$ & 4500\,K & assumed \\
q ($M_2/M_1$)   & $0.094$& $\pm0.002$\\
$\Omega_{1}=\Omega_{2}$ & 1.8854 & $\pm0.0249$\\
$T_{2}$ & 4354\,K & $\pm161$K \\
$i$     & 73.4  & $\pm7.0$\\
$L_{1}/(L_{1}+L_{2})$ (I) &$0.8929$ & $\pm0.0032$\\
$r_{1}(pole)$&$0.5546$ & $\pm0.0076$\\
$r_{1}(side)$&$0.6292$ & $\pm0.0133$\\
$r_{1}(back)$&$0.6514$ & $\pm0.0158$\\
$r_{2}(pole)$&$0.2074$ & $\pm0.0129$\\
$r_{2}(side)$&$0.2189$ & $\pm0.0161$\\
$r_{2}(back)$&$0.2980$ & $\pm0.0990$\\\hline\hline
\end{tabular}
\end{table}

   \begin{figure*}
   \centering
   \includegraphics[width=\textwidth, angle=0]{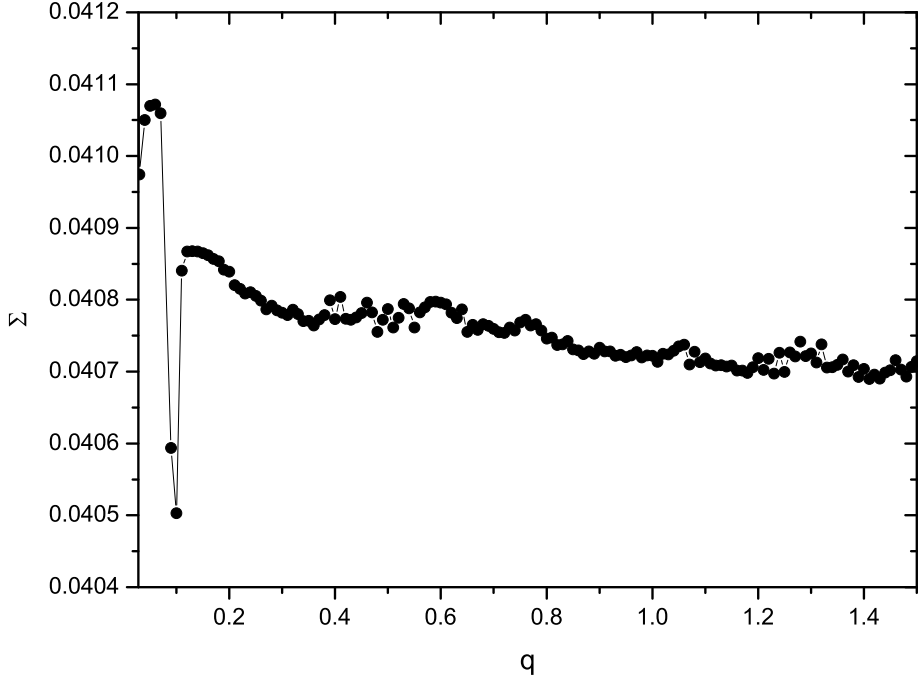}
   \caption{The relation between $\Sigma$ and q obtained based on the photometric
data in 2002. It is shown that the minimum of $\Sigma$ is at q=0.1.}
   \label{fig2}
   \end{figure*}

The q-search method was used to determine the mass ratio. We focus
on searching for photometric solutions with mass ratio from 0.03 to
1.5, and solutions were carried out for 147 values of the mass
ratio. The relation between the resulting sum $\Sigma$ of weighted
square deviations and q is plotted in Fig. 2. It is found that the
solution converged at q=0.1 with the lowest value of $\Sigma$
indicating that the theoretical light curve based on the solution is
the best one to fit the observations. Then, q was treated as an
adjustable parameter and the value of 0.1 was taken as the initial
value. Finally photometric elements were obtained and it is found
that the solution converges at q=0.094. The corresponding
photometric solutions are listed in Table 1.

As shown in Fig. 1, the light curve of V1309 Sco in 2002 displayed a
negative O'Connell effect. i.e., the light maximum following the
primary minimum is lower than the other one. The components of V1309
Sco are cool stars. The deep convective envelope along with rapid
rotation can produce a strong magnetic dynamo and solar like
magnetic activity. It is expected that dark spots should be observed
on photospheres. In the W-D method, there are four parameters for
each spot: spot center longitude ($\theta$), spot center latitude
($\phi$), spot angular radius ($r$), and spot temperature fact
($T_f$), all in units of radian. Our solution suggests that the
asymmetry of the light curves can be plausibly explained as the
presence of one dark spot on the more massive component. The
parameters of the dark spot are shown in Table 2. The theoretical
light curves is plotted in Fig. 1 as the solid line that fits the
observations well. The corresponding geometric structure at phase
0.25 is displayed in Fig. 3. The photometric solution indicates that
the temperature of the dark spot is about 350\,K lower than that of
the stellar photosphere on the more massive component star. The dark
spot covers 1.8\,\% of the total photospheric surface that is much
larger than that of a spot on the Sun (the area of sunspot is
usually less than 1\,\% of the photospheric surface of the Sun).
However, the solution of the dark spot derived with the W-D method
is definitely not unique. The spot may be composed of a group of
small spots.

\begin{table}
\caption{Parameters of the dark spot on the more massive component.}
\begin{tabular}{ll}\hline\hline
   Spot parameters & value \\\hline
   $\theta(radian)$ &   1.390 \\
   $\phi(radian)$ &   4.688 \\
   $r(radian)$ &   0.224\\
   $T_f(T_d/T_0)$ &   0.923 \\
\hline\hline
\end{tabular}
\end{table}

   \begin{figure*}
   \centering
   \includegraphics[width=\textwidth, angle=0]{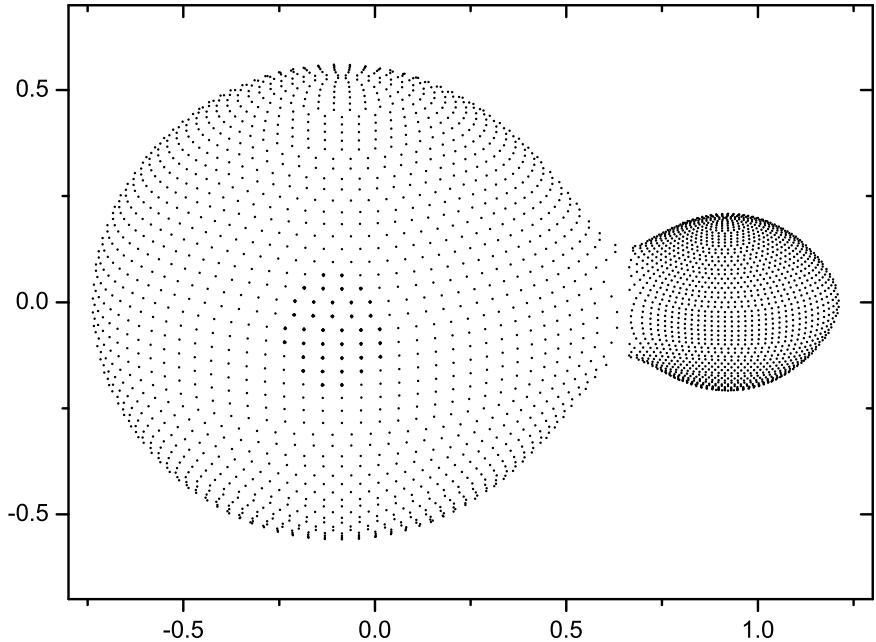}
   \caption{Geometrical structure of the progenitor of V1309 Sco at phase 0.25.}
   \label{fig3}
   \end{figure*}

\section{Discussions and conclusions}

The photometric solution of the 2002 light curve suggests that V1309
Sco is an extreme low mass ratio contact binary star
($q=M_2/M_1=0.094$) with an extremely high fill-out factor
($f=89.5(\pm40.5)\,\%$). The asymmetry of the light curve was
explained by the presence of a dark stellar spot on the more massive
component. The extremely high fill-out factor indicates that the
common convective envelope reaches the outer critical Roche lobe and
causes a great quantity of mass and angular momentum loss. This
suggests that the merging of the contact binary is driven by
dynamical mass loss from the outer Lagrange point. This is in
agreement with the rapidly decay in the orbital period (e.g., Pejcha
2014). The period decay timescale $P/\dot{P}$ was decreasing from
$\sim1000$ to $\sim170$\,years in about 6 years, which suggests that
the dynamically mass-loss rate was  increasing.

The binary progenitor of V1309 Sco consists of a giant and a
main-sequence companion (e.g., Nandez et al. 2014). The masses of
the primary and the secondary were estimated as 1.52\,$M_{\odot}$
and 0.16\,$M_{\odot}$ by Nandez et al. (2014) with a mass ratio of
$\sim0.105$. This is consistent with the present value ($q=0.094$).
According to the result obtained by Stepien (2011), the instability
resulting in the merging of both components was triggered by a
dramatic increase of the moment of inertia of the component star
when it approached the base of the red giant branch. However, our
photometric solution indicates that the merging is produced the
dynamically mass loss through the $L_2$ point.

   \begin{figure*}
   \centering
   \includegraphics[width=\textwidth, angle=0]{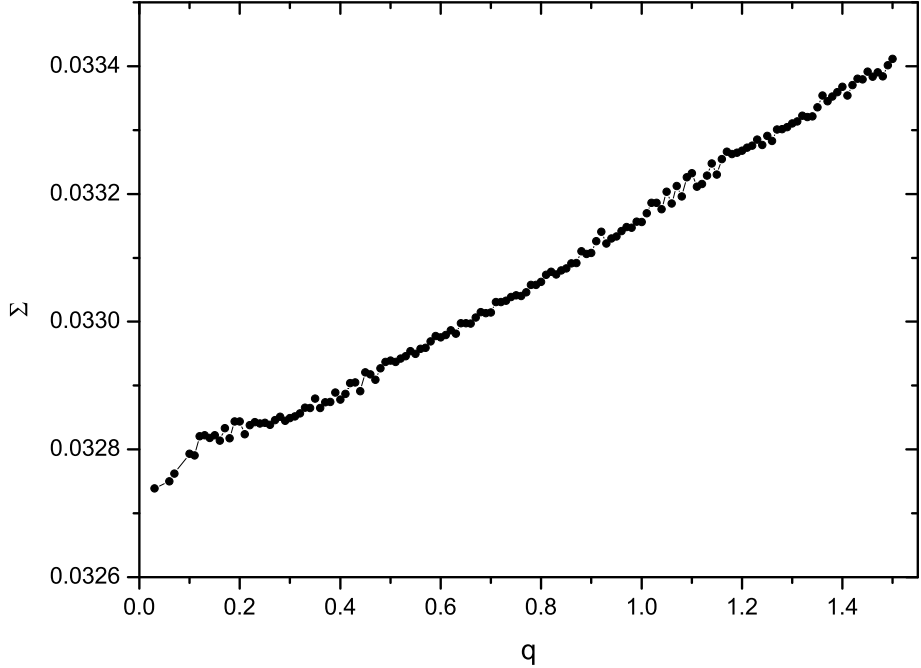}
   \caption{The relation between $\Sigma$ and q obtained based on the photometric
data in 2004. It is shown that $\Sigma$ is decreasing with the mass ratio $q$ continuously.}
   \label{fig4}
   \end{figure*}

To understand the physical properties of V1309 Sco before merging,
we also analyzed the I-band light curve in 2004 (open triangles).
The relation between $\Sigma$ and q is shown in Fig. 4. As displayed
in the figure, the value of $\Sigma$ is decreasing with the mass
ratio $q$ continuously. This indicates that no reliable solutions
can be obtained even at an extremely low mass ratio q=0.03. The
theoretical light curve in Fig. 1 (the dashed line) was calculated
by fixed q=0.03. This could be explained as the disappear of the
common convective envelope of the contact binary system suggesting
that the main-sequence companion has started to spiral in the
envelope of the giant primary and formed a real "common envelope".
This evolutionary process was discussed by Ivanova et al. (2013a,b).
The circumstellar material obscured the binary system, which may
cause the optical brightness dip before the outburst (see Fig. 1 in
the paper by Tylenda et al. (2011)).

Some low-mass-ratio and deep contact binary stars were discovered by
Qian et al. (2005a, 2011), Yang et al. (2012,
2013), Samec et al.(2011) and Zhu et al. (2005, 2011) where the fill-out factor ($f$)
is higher than 50\,\%. The orbital periods of some targets are
decreasing continuously. Some examples are GR Vir (Qian \& Yang,
2004), FG Hya (Qian \& Yang, 2005), IK Per (Zhu et al., 2005), CU
Tauri and TV Muscae (Qian et al. 2005b), and XY LMi (Qian et al.
2011). As the period is decreasing, the inner and outer critical
Roche lobes will be shrinking and thus will cause $f$ increasing.
Finally, they will merge into a single rapidly rotating star and
produce the eruption like one observed for V1309 Sco when the
convective surface reaches the outer critical Roche Lobe. However,
the pre-burst contact binary V1309 is different from those W
UMa-type stars. Those stars have their primaries on the
main-sequence stage and stay in contact for a long time, while the
primary of V1309 Sco is a giant. Continuous monitor of those stars
is very useful to understand the merge of contact binary stars.

\begin{acknowledgements}
This work is partly supported by Chinese Natural
Science Foundation (No.11133007 and No. 11325315), the Key Research
Program of the Chinese Academy of Sciences (Grant No. KGZD-EW-603),
the Science Foundation of Yunnan Province (Nos. 2012HC011 and
2013FB084), and by the Strategic Priority Research Program "The
Emergence of Cosmological Structures" of the Chinese Academy of
Sciences (No. XDB09010202).
\end{acknowledgements}

\label{lastpage}


\begin{thebibliography}{99}

\bibitem{}Ivanova, N., Justham, S., Chen, X., De Marco, O., Fryer, C. L., et al., 2013, A\&ARv 21, 59
\bibitem{}Ivanova, N., Justham, S., Avendano Nandez, J. L., Lombardi, J. C., 2013, Sciece 339, 433
\bibitem{}Mason, E., Diaz, M., Williams, R. E., Preston, G., \& Bensby, T., 2010, A\&A 516,
A10
\bibitem{}McCollum, B., Laine, S., et al., 2014, AJ 147, 11
\bibitem{}Nakano, S. 2008, IAU Circ., 8972
\bibitem{}Nandez, J. L. A., Ivanova, N., Lombardi, Jr. J. C., 2014, ApJ 786, 39
\bibitem{}Pejcha, O., 2014, ApJ 788, 22
\bibitem{}Qian, S.-B., 2001a, MNRAS 328, 635
\bibitem{}Qian, S.-B., 2001b, MNRAS 328, 914
\bibitem{}Qian, S.-B., 2002a, MNRAS 336, 1247
\bibitem{}Qian, S.-B., 2002b, A\&A 387, 903
\bibitem{}Qian, S.-B., 2003a, MNRAS 342, 1260
\bibitem{}Qian, S.-B., 2003b, A\&A 400, 649
\bibitem{}Qian, S.-B., Liu, L., Zhu, L.-Y., He, J.-J., Yang, Y.-G., Bernasconi, L., 2011, AJ 141, 151
\bibitem{}Qian, S.-B. \& Yang, Y.-G., 2004, AJ 128, 2430
\bibitem{}Qian, S.-B. \& Yang, Y.-G., 2005, MNRAS 356, 765
\bibitem{}Qian, S.-B., Yang, Y.-G., Soonthornthum, B., Zhu, L.-Y.,
He, J.-J., and Yuan, J.-Z., 2005b, AJ 130, 224
\bibitem{}Qian, S.-B., Zhu, L.-Y., Soonthornthum, B., Yuan, J.-Z.,
Yang Y.-G., and He, J.-J., 2005a, AJ 130, 1206
\bibitem{}Rudy, R. J., Lynch, D. K., Russell, R. W., et al. 2008a, IAU Circ., 8976
\bibitem{}Rudy, R. J., Lynch, D. K., Russell, R. W., et al. 2008b, IAU Circ., 8997
\bibitem{}Samec, R. G., Labadorf, C. M., Hawkins, N. C., Faulkner, D. R., Van Hamme, W., 2011, AJ 142, 117
\bibitem{}Stepien, K., 2011, A\&A 531, A18S
\bibitem{}Tylenda, R., Hajduk, M., Kami¨½ski, T., Udalski, A., Soszy¨½ski, I., Szyma¨½ski, M. K., Kubiak, M., Pietrzy¨½ski, G., Poleski, R., et al., 2011, A\&A 528, A114
\bibitem{}Tylenda, R., \& Soker, N. 2006, A\&A, 451, 223
\bibitem{}van Hamme, W., 1993, AJ 106, 2096
\bibitem{}Wilson, R. E., 1994, PASP 106, 921
\bibitem{}Wilson, R. E. \& Devinney, E. J., 1971, ApJ 166, 605
\bibitem{}Yang, Y.-G., Qian, S.-B., Soonthornthum, B., 2012, AJ 143, 122
\bibitem{}Yang, Y.-G., Qian, S.-B., Zhang, L.-Y., Dai, H.-F., Soonthornthum, B., 2013, AJ 146, 35
\bibitem{}Zhu, L.-Y. \& Qian, S.-B., 2006, MNRAS 367, 423
\bibitem{}Zhu, L.-Y. \& Qian, S.-B., 2009, AJ 138, 2002
\bibitem{}Zhu, L.-Y., Qian, S.-B., Soonthornthum, B., and Yang,
Y.-G., 2005, AJ 129, 2806
\bibitem{}Zhu, L. Y., Qian, S. B., Soonthornthum, B., He, J. J., anf Liu, L., 2011, AJ 142, 124
\bibitem{}Zhu, L.-Y., Zejda, M., Mikul\'{a}\v{s}ek, Z.; Li\v{s}ka, J., Qian, S.-B., de Villiers, S. N., 2012, AJ 144, 37
\bibitem{}Zhu, L. Y., Qian, S. B., Zola, S., Kreiner, J. M., 2009, AJ 137, 3574

\end{thebibliography}
\end{document}